\documentclass[a4paper,12pt]{article}
\usepackage{amsfonts}
\usepackage{latexsym}
\usepackage{amssymb}
\usepackage{amsmath}
\usepackage{color}
\date{}

\begin{document}

\title{Non-Equilibrium Thermodynamics and Stochasticity\\
A Phenomenological Look on Jarzynki's Equality}
\author{W. Muschik\footnote{Corresponding author: muschik@physik.tu-berlin.de}
\\
Institut f\"ur Theoretische Physik\\
Technische Universit\"at Berlin\\
Hardenbergstr. 36\\D-10623 BERLIN,  Germany}
\maketitle

           \newcommand{\R}[1]{(\ref{#1})}
           \newcommand{\C}[1]{\cite{#1}}

            \newcommand{\mvec}[1]{\mbox{\boldmath{$#1$}}}
            \newcommand{\x}{(\!\mvec{x}, t)}
            \newcommand{\m}{\mvec{m}}
            \newcommand{\F}{{\cal F}}
            \newcommand{\n}{\mvec{n}}
            \newcommand{\argm}{(\m ,\mvec{x}, t)}
            \newcommand{\argn}{(\n ,\mvec{x}, t)}
            \newcommand{\T}[1]{\widetilde{#1}}
            \newcommand{\U}[1]{\underline{#1}}
            \newcommand{\ov}[1]{\overline{#1}}
            \newcommand{\X}{\!\mvec{X} (\cdot)}
            \newcommand{\cd}{(\cdot)}
            \newcommand{\Q}{\mbox{\bf Q}}
            \newcommand{\p}{\partial_t}
            \newcommand{\z}{\!\mvec{z}}
            \newcommand{\bu}{\!\mvec{u}}
            \newcommand{\rr}{\!\mvec{r}}
            \newcommand{\w}{\!\mvec{w}}
            \newcommand{\g}{\!\mvec{g}}
            \newcommand{\D}{I\!\!D}
            \newcommand{\se}[1]{_{\mvec{;}#1}}
            \newcommand{\sek}[1]{_{\mvec{;}#1]}}            
            \newcommand{\seb}[1]{_{\mvec{;}#1)}}            
            \newcommand{\ko}[1]{_{\mvec{,}#1}}
            \newcommand{\ab}[1]{_{\mvec{|}#1}}
            \newcommand{\abb}[1]{_{\mvec{||}#1}}
            \newcommand{\td}{{^{\bullet}}}
            \newcommand{\eq}{{_{eq}}}
            \newcommand{\eqo}{{^{eq}}}
            \newcommand{\f}{\varphi}
            \newcommand{\ro}{\varrho}
            \newcommand{\dm}{\diamond}
            \newcommand{\seq}{\stackrel{_\bullet}{=}}
            \newcommand{\st}[2]{\stackrel{_#1}{#2}}
            \newcommand{\om}{\Omega}
            \newcommand{\emp}{\emptyset}
            \newcommand{\bt}{\bowtie}
            \newcommand{\btu}{\boxdot}
\newcommand{\Section}[1]{\section{\mbox{}\hspace{-.6cm}.\hspace{.4cm}#1}}
\newcommand{\Subsection}[1]{\subsection{\mbox{}\hspace{-.6cm}.\hspace{.4cm}
\em #1}}

\vspace{.8cm}
\noindent
{\bf Abstract}
The theory of phenomenological Non-equilibrium Thermodynamics
is extended by includimg stochastic processes in order to account for recently derived
thermodynamical relations such as the Jarzynski equality.
Four phenomenological axioms are postulated resulting in a phenomenological interpretation of Jarzynski's equality.  Especially, considering the class of Jarzynski processes Jarzynski's equality follows from the axiom that the statistical 
average of the exponential work is protocol independent.
\vspace{.5cm}\newline

\section{Introduction}

Among many branches of thermodynamics {\bf --}such as Thermodynamics of
Irreversible
Processes, Rational Thermodynamics, Extented Thermodynamics, Endoreversible
Thermodynamics, Finite Time Thermodynamics, Quantum Thermodynamics, Mesoscopic
Theory, GENERIC \C{MU1}{\bf --}
Stochastic Thermodynamics is another branch which introduces probabilities into
the thermodynamical description \C{SEIF}. But in contrast to the above
mentioned branches,
Stochastic Thermodynamics allows processes of negative process entropy,
some times called "violations" of the Second Law.
This item does not have any consequences on the phenomenological level because
the negative process entropies vanish by establishing mean values using the
introduced probabilities. Consequently it is obvious, how to obtain the
phenomenological level
in Stochastic Thermodynamics, but how vice-versa to incorporate stochastic
processes with their
probabilities and process entropies into phenomenological Non-equilibrium
Thermodynamics is an open question which is investigated in this paper.
This is not done in full generality, but only for a special process class
{\bf--}called the Jarzynski process class{\bf--} because we want to clarify the
status of the integral fluctuation theorem
{\bf--}the Jarzynski equality \cite{1}{\bf--} in the framework of
Non-equilibrium Thermodynamics.
\vspace{.3cm}\newline
The paper is organized as follows: After having introduced the Jarzynski
process, a sketch of Non-equilibrium Thermodynamics of
discrete systems is given for introducing the items which later on are
needed. The Jarzynski process of phenomenological Non-equilibrium
Thermodynamics is replaced by a set of stochastic processes which generate probability
densities defined on the process work as a stochastic variable. These stochastic
processes are decomposed into regular and non-regular processes distinguished by their
process entropy production: regular ones have positive process entropy
production, whereas that of the non-regular processes is negative.
The mean values of process dissipation and work are considered with respect
to the phenomenological Second Law. Two phenomenological axioms establish
Jarzynski's equality. The reversible case and equilibrium are shortly
discussed. A summary and a discussion finish the paper.

\section{The Jarzynski Process\label{JP}}

We consider a discrete system\footnote{A so-called Schottky
system \cite{SS}: a ``box'' interacting with its environment which is not to be confused with discrete 
systems in Stochastic Thermodynamics which evolve on a discrete state space \cite{SEIF}.}
which interacts with its environment by power and
heat exchange performing a Jarzynski process. Such a process
begins at time $t=0$ in a fixed
equilibrium state $A^{eq}$ whose work variable (Jarzynski's switching
parameter) is $\lambda = 0$ and arrives at time $t=\tau$ in a
non-equilibrium state\footnote{A so-called {\em non-thermal state}.} $C^{neq}$
with the work variable $\lambda = 1$. During the time 
$0\leq t\leq\tau$, the system exchanges irreversibly work and
heat with its controlling equilibrium environment\footnote{A heat reservoir.}
of constant thermostatic temperature $T^*$. For times
$\tau< t\leq d$ the thermal contact with the controlling
heat reservoir maintains, but the switching parameter is fixed
at $\lambda = 1$, that means, no additional work, but heat is
exchanged between the system and the reservoir during this time
interval. Consequently, the final equilibrium state $B^{eq}$
has the same work variable $\lambda = 1$ as $C^{neq}$,
and according to the process control by the heat reservoir
its temperature is $T^*$. Because the state space of
the system in consideration is given by the work variable $\lambda$
and by the contact temperature $\Theta$\ \footnote{The contact
temperature is a non-equilibrium analogue of the
thermostatic temperature. More details in sect.\ref{CT} \cite{5,6,9}.},
we can represent the Jarzynski process as follows
\begin{eqnarray}\label{J1}
{\cal J}_\lambda:\quad &t=0:&\ A^{eq}(\lambda=0,T^*)
\quad\longrightarrow\quad [\st{\td}{\lambda}(t),\st{\td}{Q}(t)]
\quad\longrightarrow
\\ \label{J2}\longrightarrow\quad
&t=\tau:&\ \ C^{neq}(\lambda=1,\Theta).\quad\longrightarrow\quad
[\st{\td}{\lambda}=0,\st{\td}{Q}(t)]\quad\longrightarrow
\\ \label{J3}\longrightarrow\quad
&t=d:&\ \ B^{eq}(\lambda=1,T^*).
\end{eqnarray}
Here, the time-dependent work variable $\lambda(t)$ {\bf--}the protocol{\bf--}
is arbitrary, but controlled, whereas the heat exchange
$\st{\td}{Q}(t)$ is uncontrolled and depends on the given protocol.
\vspace{.3cm}\newline
The interpretation of the work which is done during the transfer
from $\lambda =0$ to $\lambda =1$ {\bf--}during the protocol{\bf--} is different in
Non-equilibrium and Stochastic Thermodynamics: in Non-equilibrium
Thermodynamics the work is a phenomenological variable, whereas in Stochastic
Thermodynamics the work is introduced as a stochastic variable $w$ generating a
propability distribution $p_\lambda(w)$, if identical protocols $\lambda(t)$ are
performed. Considering a Jarzynski process \R{J1} to \R{J3}
with a heat reservoir at inverse (thermostatic) temperature $\beta$,
the probability distribution
is not arbitrary, but constrained by the relation which is known as Jarzynski equality
\begin{equation}\label{eq Jarzynski equality}
 \int_{-\infty}^\infty p_\lambda(w) \exp(-\beta w)dw = \exp(-\beta\Delta^{AB} F),
\end{equation}
where $\Delta^{AB} F$ denotes the difference of the
\emph{equilibrium} free energy of the system between the 
states $A^{eq}$ and $B^{eq}$. Note 
that ~(\ref{eq Jarzynski equality}) remains valid even if we only look at 
changes of the system from $A^{eq}$ to $C^{neq}$ because there is no work
performed on the system from $C^{neq}$ to 
$B^{eq}$ and $p_\lambda(w)$ remains unchanged during that part. Hence,
~(\ref{eq Jarzynski equality}) teaches that non-equilibrium stochastic
fluctuations contain valuable information about equilibrium quantities. 
\vspace{.3cm}\newline
Jarzynski's equality and was originally discovered in 
1997~\cite{1, JarzynskiPRE1997}. Since then it was derived under many circumstances, for instance, it follows from 
the detailed (Crooks) fluctuation theorem~\cite{CrooksJSP1998, CrooksPRE1999, CrooksPRE2000}, it is also valid in the 
strong coupling regime~\cite{JarzynskiJSM2004} or for quantum systems~\cite{TasakiArXiv2000, MukamelPRL2003, 
CampisiTalknerHaenggiPRL2009}. Furthermore, early experimental confirmations can be found in~\cite{TrepagnierEtAlPNAS2004, CollinEtAlNature2005, BustamanteLiphardtRitortPhysToday2005} and a recent review 
is given in~\cite{JarzynskiReview2011}. In the next section we will look at the Jarzynski process from a purely 
phenomenological perspective (without probabilities and stochasticity) before we then ask how to incorporate the Jarzynski equality on a phenomenological level ?

\section{A Thermodynamical Sketch}
\subsection{Basic phenomenological variables}

The state of a discrete thermodynamical system is described by
basic variables. Kind and number of these variables depend on the nature
of the system under consideration and on the process going on in that system.
The number of variables in non-equilibrium is clearly greater than in
equilibrium. Therefore, equilibrium needs a minimal number of basic variables
spanning the so-called equilibrium sub-space. According to a special
formulation of the Zeroth Law \cite{2}, the equilibrium variables of a thermally
homogeneous\footnote{A system without adiabatic partitions.}
system are
\begin{equation}\label{1}    
\mvec{z}_{eq}\ =\ (U,\mvec{a},\mvec{n})\qquad\mbox{or}\qquad
\mvec{z}_{eq}\ =\ (T,\mvec{a},\mvec{n})
\end{equation}
(internal energy U, work variables \mvec{a}, mol numbers \mvec{n},
thermostatic temperature T). In equilibrium,
there exists an one-to-one mapping between the
internal energy of the system and its thermostatic temperature
$U\leftrightarrow T$.
\vspace{.3cm}\newline
More basic variables than in equilibrium are needed in non-equilibrium
\begin{equation}\label{2}
\mvec{z}\ =\ (U,\mvec{a},\mvec{n},\mvec{z}_{neq}).
\end{equation}
The set of the non-equilibrium variables {$\mvec{z}_{neq}$}
depends on the nature of the system in
consideration: e.g. the orientation of needle-shaped molecules may be an
example in the case of complex materials, time derivatives of the equilibrium
variables and dissipative fluxes are other examples. Here with regard to Jarzynski
processes, we introduce the contact temperature $\Theta$ of the system as one
non-equilibrium variable \cite{5,6,9}. Other basic non-equilibrium
variables {\bf--}e.g. the internal variables{\bf--} are marked by a
place-holder $\mvec{\xi}$. Consequently, the non-equilibrium variables are
\begin{equation}\label{2a}
\mvec{z}_{neq}\ =\ (\Theta, \mvec{\xi}),
\end{equation}
and the basic variables spanning the state space of the system are
\begin{equation}\label{7}
\mvec{z}\ =\ (U, \mvec{a}, \mvec{n}, \Theta, \mvec{\xi}).
\end{equation}
The non-equilibrium contact temperature $\Theta$ is independent of the other variables of
the state space, especially independent of the internal energy \cite{10}.

\subsection{Non-equilibrium entropy, First Law}

A non-equilibrium entropy is a state function on the non-equilibrium state space \R{7}.
The time rate of this non-equilibrium entropy is an analogue to Gibbs'
fundamental equation \cite{9}
\begin{equation}\label{6}
\st{\td}{S}\ :=\ \frac{1}{\Theta}\st{\td}{U} -  \frac{\mvec{A}}{\Theta}\cdot
\st{\td}{\mvec{a}} - \frac{\mvec{\mu}}{\Theta}\st{\td}{\mvec{n}} +
\alpha\st{\td}{\Theta} + \mvec{\beta}\cdot\mvec{\st{\td}{\xi}}
\end{equation}
The conjugate quantities to the state space variables are: the reciprocal contact
temperature $1/\Theta$, the generalized forces $\mvec{A}$ over the contact
temperature, the chemi\-cal potentials $\mvec{\mu}$ over the contact
temperature and  the conjugate quantities $\alpha$ and $\mvec{\beta}$ related to
the contact temperature and to the internal variables\footnote{If the contact
temperature $\Theta$ in \R{6} is replaced by the thermostatic
temperature $T^*$ of the controlling heat reservoir, the expression
\R{6} looses its property as state function because $T^*$ belogs to the controllimg
heat reservoir and is therefore not a state variable. 
In equilibrium, $\Theta$ is replaced by the thermostatic temperature $T$ of the system.}.
\vspace{.3cm}\newline
Introducing the molar enthalpy $\mvec{h}$ and the external change of mol
numbers $\st{\td}{\mvec{n}}\!{^e}$ by the material exchange between system and
environment \cite{8}, the First Law and the power $\st{\td}{W}$ are 
\begin{equation}\label{8}
\st{\td}{U}\ =\ \st{\td}{Q} + \st{\td}{W} + \mvec{h}\cdot\st{\td}{\mvec{n}}{^e}
,\qquad \st{\td}{W}\ :=\ \mvec{A}\cdot\st{\td}{\mvec{a}}.
\end{equation}
Approaching Jarzynski processes, we consider here closed discrete systems
without chemical reactions
\begin{equation}\label{9}
\st{\td}{\mvec{n}}\ \doteq\ \mvec{0},\qquad\st{\td}{\mvec{n}}{^e}\ \doteq\
\mvec{0}.
\end{equation}
By taking (\ref{8}) and (\ref{9}) into account, (\ref{6}) results for closed
systems in 
\begin{equation}\label{N30}
\st{\td}{S}\ =\ \frac{1}{\Theta}\st{\td}{Q} +
\alpha\st{\td}{\Theta} +\mvec{\beta}\cdot\mvec{\st{\td}{\xi}} .
\end{equation}
This is the usual expression of decomposing the entropy time rate in phenomenological
Non-equilibrium Thermodynamics~\C{GM}:
$\st{\td}{Q}/\Theta$ is the entropy flux and
$\alpha\st{\td}{\Theta} +\mvec{\beta}\cdot\mvec{\st{\td}{\xi}}$ the entropy
production. Closing the system enforces vanishing of the entropy flux.

\subsection{Contact temperature, free energy and work\label{CT}}

We consider a closed non-equilibrium system which is in contact with a heat
reservoir of thermostatic temperature $T^*$. The heat exchange between them
is $\st{\td}{Q}$. We now define the contact temperature $\Theta$ of the
non-equilibrium system \cite{5,6} by the inequality
\begin{equation}\label{N30a}
\left(\frac{1}{\Theta}-\frac{1}{T*}\right)\st{\td}{Q}\ \geq\ 0.
\end{equation}
This ``defining inequality of the contact temperature'' states that $\Theta = T^*$
 if and only if $\dot Q = 0$. 
\vspace{.3cm}\newline
Taking (\ref{N30}), (\ref{8}), (\ref{9}) and \R{N30a} into account, we obtain
\begin{equation}\label{N30b}
\st{\td}{S} -\alpha{\st{\td}\Theta}-\mvec{\beta}\cdot\mvec{\st{\td}{\xi}}\ \geq\ 
\frac{1}{T^*}\Big(\st{\td}{U}-\st{\td}{W}\Big),
\end{equation}
an inequality which stems from introducing the contact temperature by \R{N30a}
and which is independent of the Second Law.
Here, $T^*$ is the constant thermostatic temperature of a controlling heat
reservoir, such one {which appears} in the Jarzynski process.
\vspace{.3cm}\newline
Next, we define the state function of
non-equilibrium free energy as\footnote{This definition is in contrast to 
definitions used in Stochastic Thermodynamics~\cite{GaveauSchulmanPLA1997, CrooksPRE2007, EspositoVandenBroeckEPL2011} 
where the contact temperature is not used and is replaced by the
thermostatic temperature $T^*$ of the controlling reservoir. Thus, $F$ becomes
the free energy in the equilibrium state $B^{eq}$ \R{J3}, whereas \R{10} refers to $C^{neq}$.}
\begin{equation}\label{10}
F(\mvec{a},\Theta,\mvec{\xi})\ :=\ U - \Theta S
\quad\Longrightarrow\quad
\st{\td}{F}\ =\ \st{\td}{U} - (\Theta S)^\td.
\end{equation}
Inserting (\ref{10}) into (\ref{N30b}) results in
\begin{equation}\label{N30c}
\st{\td}{S} -\alpha{\st{\td}\Theta}-\mvec{\beta}\cdot\mvec{\st{\td}{\xi}}\ \geq\ 
\frac{1}{T^*}\Big(\st{\td}{F} +(\Theta S)^\td  -\st{\td}{W}\Big).
\end{equation}
Integration along a Jarzynski process ${\cal J}_\lambda:
A^{eq}\longrightarrow B^{eq}$ yields
\begin{equation}\label{N30d}
S_B - S_A -{\cal J}_\lambda\!\!\!\int_A^B(\alpha{\st{\td}\Theta}+\mvec{\beta}\cdot\mvec{\st{\td}{\xi}})dt\ \geq\
\frac{1}{T^*}\Big(\Delta^{AB}F +\Theta_BS_B - \Theta_AS_A - W^{AB}\Big).
\end{equation}
According to (\ref{J1}) and (\ref{J3}),
\begin{equation}\label{N30e}
\Theta_A\ =\ \Theta_B\ =\ T^* 
\end{equation}
is valid\footnote{An integration only between $A^{eq}$ and $C^{neq}$ would not
result in \R{N30e} because of $\Theta_C\neq T^*$.}, 
and we obtain an inequality valid along Jarzynski processes
\begin{equation}\label{N30f}
{\cal J}_\lambda\!\!\!\int_A^B(\alpha{\st{\td}\Theta}+\mvec{\beta}\cdot\mvec{\st{\td}{\xi}} )dt\ =:\
\Sigma^{AB}\ \leq\ \frac{1}{T^*}\Big(W^{AB} - \Delta^{AB}F\Big)\
=:\ {\frac{D^{AB}}{T^*}}
\end{equation}
which stems { as \R{N30b}} from the defining inequality of the contact temperature (\ref{N30a}). 
According to the decomposition of the entropy rate \R{N30}, the bracket in
\R{N30f}$_1$ is the entropy production, so that 
$\Sigma^{AB}$ becomes the process entropy production 
between $A^{eq}$ and $B^{eq}$, and
$D^{AB}$  is the process dissipation which
is in generally greater than $T^*\Sigma^{AB}$. 
Note that $D^{AB}/T^*$ is also sometimes called the entropy production because 
it coincides with the entropy increase of system \emph{and} 
bath~\cite{1, SEIF, JarzynskiReview2011}.
\vspace{.3cm}\newline
The process work $W^{AB}$ along the Jarzynski process ${\cal J}_\lambda$ is
related to the given protocol $\lambda(t)$ according to \R{J1} and to the generalized
forces $ L({\lambda,\Theta})$
\begin{equation}\label{J4}
W^{AB}_\lambda\ =\ {\cal J}_\lambda\!\!\!\int_A^B L({\lambda,\Theta})
\st{\td}{\lambda}(t)dt.
\end{equation}
This quantity will become a key position when stochastic processes
are introduced below. We now take the Second Law into account.

\subsection{Second Law}

According to the Second Law, the entropy production
\begin{equation}\label{N31}
\alpha\st{\td}{\Theta} +\mvec{\beta}\cdot\mvec{\st{\td}{\xi}}\ \geq\ 0,
\end{equation}
is not negative in phenomenological Non-equilibrium Thermodynamics \C{GM}.
Thus, (\ref{N30}) and (\ref{N30f})$_1$ become with (\ref{N30a}) and
(\ref{N31})
\begin{equation}\label{N32}
\st{\td}{S}\ \geq\ \frac{1}{\Theta}\st{\td}{Q}\ \geq\ 
\frac{1}{T^*}\st{\td}{Q},\qquad 0\ \leq\ \Sigma^{AB}\ \leq\
\frac{D^{AB}}{T^*}.
\end{equation}
From \R{N30f}$_2$ and \R{N32}$_3$ follows the well known fact which is also
valid for Jarzynski processes
\begin{equation}\label{aJ4}
W^{AB}\ \geq\ \Delta^{AB}F.
\vspace{.3cm}\end{equation}
The reversible case is defined by
\begin{equation}\label{J4a}
(\alpha\st{\td}{\Theta} + \mvec{\beta}\cdot\mvec{\st{\td}{\xi}})^{rev}
\ \equiv\ 0,\qquad \mbox{and}\quad
\Theta\ \equiv\ T^*\ =\ \mbox{const}.
\end{equation}
Consequently,  we obtain according to \R{N30f} and \R{J4a}$_1$  for the reversible
case
\begin{equation}\label{J4b}
\Sigma^{AB}_{rev}\ =\ D^{AB}_{rev}\ =\ 0\quad\longrightarrow\quad
W^{AB}_{rev}\ =\ \Delta^{AB}F.
\vspace{.3cm}\end{equation}
This sketch outlines the tools which we need in the sequel.

\section{Introducing Stochasticity}

In contrast to macroscopic systems, the behaviour of small (mesoscopic) systems is inherently stochastic:
for describing them, stochastic variables have to be used generating
probability distributions. Although the average behaviour of stochastic systems
still obeys the phenomenological laws of thermodynamics, there is much more to
discover as Stochastic Thermodynamics lets suppose.
Here, for the special exampe of Jarzynki processes, we are interested in a ``top-down'' approach, i.e., we ask how 
to modify phenomenological Non-equilibirum Thermodynamics in order to account for fluctuations and stochasticity. 
This is in contrast to Stochastic Thermodynamics, which follows rather from a
``bottom-up'' approach by relying on microscopically derived equations of motion
(e.g., Langevin or master equations)~\cite{SEIF}.

\subsection{Process work as a stochastic variable}

We consider one of the numerous protocols $\lambda(t)$ performing
a Jarzynski process. According to \R{J4}, the work $W^{AB}_\lambda$ is
required. Whenever the same protocol is used in Non-equilibrium
Thermodynamics, the same work is required for performing the
corresponding Jarzynski process. This situation is
totally different for {\em stochastic systems}: several experiments,
all performed with the same given protocol $\lambda(t)$ require
several different works for performing the Jarzynski process
with the result, that \R{J4} cannot hold true for stochastic systems.
Consequently, we postulate that the process work is a stochastic quantity.

\subsubsection{The first basic axiom}

$\blacksquare${\bf First\,Basic\,Axiom:}
The process work along a Jarzynski process is a stochastic variable
given by a stochastic equation 
\begin{equation}\label{J5}
{\cal W}^{AB}_\lambda\ =\  {\cal J}_\lambda\!\!\!\int_A^B {\cal L}({\lambda,\Theta})
\st{\td}{\lambda}(t)dt\ \in\ \mathbb{R}\ \longrightarrow\
p_\lambda({\cal W}^{AB}_\lambda)
\end{equation}
replacing \R{J4}. Performing the same protocol numerously, the values of the
process works generate a probability distribution function $p_\lambda$ on
${\cal W}^{AB}_\lambda$.
\hspace{2.4cm}$\blacksquare$
\vspace{.3cm}\newline
The same protocol $\lambda(t)$ generates by the stochastic properties
of the material {\bf--}introduced by the stochastic mapping ${\cal L}(\lambda,\Theta)$
for the generalized forces \C{OST}{\bf--} different process works
${\cal W}^{AB}_\lambda$
which all together implement a probability distribution function 
$p_\lambda({\cal W}^{AB}_\lambda)$
on the stochastic variable of the process work. Consequently,
in the framework of Non-equilibrium Thermodynamics, the probability
distribution function $p_\lambda({\cal W}^{AB}_\lambda)$ is a measurable quantity
which can be found out by performing a sufficently high number of
Jarzynski processes always using the same protocol $\lambda(t)$.
Not only the work becomes a stochastic quantity but also related quantities such
as heat, entropy and entropy production.

\subsubsection{Jarzynski process class}

We can suppose, that by replacing the phenomenological quantities of
Non-equi\-li\-brium
Thermodynamics by stochastic ones, some results of the phenomenological
theory will change: Stochastic and Non-equilibrium Thermodynamics are
different to each other, and the following question arises: What are the 
phenomenological conditions under which Non-equilibrium Thermodynamics turns
out to be a special case of Stochastic Thermodynamics ?
For instance, it is easy to see that, taking Jarzynski's inequality 
into account, processes of negative process dissipation appear in Stochastic
Thermodynamics,
a fact which is strictly forbidden in Non-equilibrium Thermodynamics.  
\vspace{.3cm}\newline
Up to now, we considered one arbitrary, but fixed protocol belonging to a special
Jarzynski process. According to \R{J1}$_3$ and \R{J2}$_3$, Jarzynski processes
can be performed with several different protocols.
All these protocols together form the {\em (stochastic) Jarzynski process class}
\begin{equation}\label{J5a}
\{{\cal J}_\lambda\}\ :=\ \Big\{\wedge\lambda:\
\lambda(t)\in {\cal J}_\lambda,\ p_\lambda({\cal W}^{AB}_\lambda)\Big\}.
\end{equation}
The introduction of this process class allows to derive connections
between the probability densities of different protocols in the sequel.

\subsection{Exponential mean process work\label{EMPW}}

Approaching Jarzynski's equality, we start out with Jensen's
inequality\footnote{Jensen's inequality 
states that for any convex function $f$ and random variable $X$ we have 
$\mathbb{E}[f(X)] \ge f[\mathbb{E}(X)]$ where $\mathbb{E}[...]$ denotes an expectation value. Since 
$f(x) = e^{-\beta x}$ is convex, Eq.~(\ref{J8}) follows.} 
\begin{eqnarray}\label{J8}
\int p(x)\exp(-\beta x)dx\ \geq\ \exp\Big(-\beta\int p(x)xdx\Big),
\\ \label{J8a}
\int p(x)dx\ =\ 1,\quad p(x)\ \geq\ 0,\quad\beta\ >\ 0,
\end{eqnarray}
and we identify $x$ with the stochastic process work, and the
probability function $p_\lambda(x)$ belongs to an arbitrary protocol 
$\lambda(t)$ 
\begin{equation}\label{J9}
x\ \equiv\ {\cal W}^{AB}_\lambda,\qquad
\int p_\lambda(x)x dx\ =:\ W^{AB}_\lambda.
\end{equation}
Here, the phenomenological process work $W^{AB}_\lambda$ is introduced
as the mean value over all stochastic process works ${\cal W}^{AB}_\lambda$.
According to this setting, Jensen's inequality \R{J8} results in 
\begin{equation}\label{J10}
\int p_\lambda({\cal W}^{AB}_\lambda)
\exp(-\beta {\cal W}^{AB}_\lambda)d{\cal W}^{AB}_\lambda\
\geq\ \exp\Big(-\beta W^{AB}_\lambda\Big).
\end{equation}
As already mentioned, the theoretical concept of process work is
different in Stochastic and Non-equilibrium thermodynamics: according to \R{J9}, we
have to distinguish between stochastic and phenomenological work:
${\cal W}^{AB}_\lambda\neq W^{AB}_\lambda$ \C{OST}.
\vspace{.3cm}\newline
Applying the mean value theorem on the lhs of \R{J10}, we obtain
\begin{equation}\label{bJ10}
\int p_\lambda(x)\exp(-\beta x)dx\ =\ \exp(-\beta M_\lambda)\
\geq\ \exp(-\beta W^{AB}_\lambda)
\end{equation}
Here, $M_\lambda$ is the {\em exponential mean process work} which 
is different from the phenomenological one according to \R{bJ10} 
\begin{equation}\label{cJ10}
M_\lambda\ \leq\ W_\lambda^{AB}.
\end{equation}
We now consider the exponential mean process work in two special cases
of the probability distribution: the non-stochastic case and the
reversible one.

\subsubsection{The non-stochastic case}

If in every repetition of the same protocol
we measure the same work value $W^{ABnst}_\lambda$, 
we refer to this case as ``non-stochastic'' and the corresponding probability density is 
\begin{equation}\label{cJ10a}
p^{nst}_\lambda(x) = \delta\left(x-W^{ABnst}_\lambda\right).	
\end{equation}
Then, according to \R{bJ10} we obtain
\begin{equation}\label{cJ10b}
\exp(-\beta W^{ABnst}_\lambda)\ =\ \exp(-\beta M^{nst}_\lambda)\
\geq\ \exp(-\beta W^{ABnst}_\lambda),
\end{equation}
resulting in
\begin{equation}\label{cJ10c}
M^{nst}_\lambda\ =\ W^{ABnst}_\lambda\ \geq\ \Delta^{AB}F,
\end{equation}
because the phenomenological process work obeys the Second Law \R{aJ4} also for
non-stochastic processes.

\subsubsection{The reversible case}

According to \R{J4b}$_2$, the reversible case is defined by
\begin{equation}\label{cJ10d}
W^{ABrev}_\lambda\ =\ \Delta^{AB}F.
\end{equation}
Thus, \R{bJ10} yields
\begin{equation}\label{bJ10e}
\int p^{rev}_\lambda(x)\exp(-\beta x)dx\ =\ \exp(-\beta M^{rev}_\lambda)\
\geq\ \exp(-\beta\Delta^{AB}F)
\end{equation}
resulting in $M^{rev}_\lambda\ \leq\ \Delta^{AB}F$. 
In comparison with (\ref{cJ10c}) and (\ref{cJ10d}), we obtain
the chain of inequalities
\begin{equation}\label{bJ10g}
M^{rev}_\lambda\ \leq\ \Delta^{AB}F\ =\ W^{ABrev}_\lambda
\leq\ M^{nst}_\lambda\ =\ W^{ABnst}_\lambda.
\end{equation}
Consequently, the exponential mean process work $M_\lambda$ is
process-dependent $\bf-$reversible or non-stochastic$\bf-$ for the
present\footnote{This process dependence vanishes by introducing Jarzynski's
equality in sect.\ref{JE}, giving rise to the second basic axiom in sect.\ref{SBA}.}.

\subsection{Regular and non-regular processes}

The stochastic process work \R{J5} takes values which can be
greater or smaller than the free energy difference $\Delta^{AB}F$.
Consequently, the integral in \R{bJ10} can be decomposed into two parts
\begin{equation}\label{fJ10}
\int_{x\geq\Delta} p_\lambda(x)\exp(-\beta x)dx
+\int_{x <\Delta} p_\lambda(x)\exp(-\beta x)dx\
=\ \exp(-\beta M_\lambda).
\end{equation}
We now denote processes with ${\cal W}^{AB}_\lambda\geq\Delta^{AB}F$
as {\em regular processes} and such with ${\cal W}^{AB}_\lambda<\Delta^{AB}F$ as
{\em non-regular} ones. Consequently, the first integral of \R{fJ10} runs over the regular
processes, whereas the second one runs over the non-regular processes.
\vspace{.3cm}\newline
Application of the mean value theorem to the lhs of \R{fJ10} is possible and results
by use of \R{bJ10} in
\begin{eqnarray}\label{gJ10}
\exp(-\beta M^+_\lambda)P^+_\lambda\ 
+\ \exp(-\beta M^-_\lambda)P^-_\lambda\ 
=\ \exp(-\beta M_\lambda)\ \geq\ \exp(-\beta W^{AB}_\lambda),
\\ \label{iJ10}
P^+_\lambda\ :=\ \int_{x\geq\Delta} p_\lambda(x)dx,\qquad
P^-_\lambda\ :=\ \int_{x <\Delta} p_\lambda(x)dx,\qquad
P^+_\lambda + P^-_\lambda\ =\ 1.
\end{eqnarray}
By construction, we obtain
\begin{equation}\label{hJ10}
 M^-_\lambda < \Delta^{AB}F ~{\le}~  M^+_\lambda.
\end{equation}
More specifically, the exponential mean process works $M^+_\lambda$ and $M^-_\lambda$ of the regular and the 
non-regular processes depend on the precise form of the probability densities.
According to \R{gJ10}, \R{fJ10} and \R{iJ10}, we obtain
\begin{eqnarray}\label{hJ10a}
M^+_\lambda\ =\ \frac{1}{\beta}\Big[\ln\int_{x\geq\Delta} p_\lambda(x)dx-
\ln\int_{x\geq\Delta} p_\lambda(x)\exp(-\beta x)dx\Big],
\\ \label{hJ10b}
M^-_\lambda\ =\ \frac{1}{\beta}\Big[\ln\int_{x<\Delta} p_\lambda(x)dx-
\ln\int_{x<\Delta} p_\lambda(x)\exp(-\beta x)dx\Big].
\end{eqnarray}
Together with \R{hJ10} and \R{iJ10}$_{1,2}$, 
this yields after a short algebraic manipulation
\begin{equation}\label{hJ10d}
\frac{P^-_\lambda}{P^+_\lambda}\ {<}\ 
\frac{\int_{x<\Delta} p_\lambda(x)\exp(-\beta x)dx}{\int_{x\geq\Delta} p_\lambda(x)\exp(-\beta x)dx}.  
\vspace{.3cm}\end{equation}
With the help of (\ref{gJ10}) and the normalization condition
{\R{iJ10}$_3$, we can solve for $P^\pm_\lambda$. This results in
\begin{eqnarray}\label{nJ10}
P^+_\lambda\ =\ 
\frac{\exp(-\beta M^-_\lambda)-\exp(-\beta M_\lambda)}
{\exp(-\beta M^-_\lambda)-\exp(-\beta M^+_\lambda)},
\\ \label{nJ10a}
P^-_\lambda\ =\ \frac{\exp(-\beta M_\lambda)-\exp(-\beta M^+_\lambda)}
{\exp(-\beta M^-_\lambda)-\exp(-\beta M^+_\lambda)}.
\end{eqnarray}
From the positivity of $P^\pm_\lambda$ we further obtain 
\begin{equation}\label{nJ10c}
M^-_\lambda \leq M_\lambda,\quad M_\lambda\leq M^+_\lambda.
\end{equation}

\subsection{Mean process work}

Up to now, the exponential mean process work was considered in sect.\ref{EMPW}.  
Let us {now} go one step back and look at the mean value of the stochastic work itself.
Starting out with \R{J9} and decomposing
its lhs according to \R{iJ10}, we obtain by use of the mean value theorem and
\R{aJ4}
\begin{equation}\label{I3a}
W^+_\lambda P^+_\lambda + W^-_\lambda P^-_\lambda\ =\ W^{AB}_\lambda\
\geq\ \Delta^{AB}F.
\end{equation}
According to the decomposition \R{iJ10}, we have for the mean values of the process
works belonging to the regular (+) and non-regular (-) processes
\begin{equation}\label{I3b}
W^-_\lambda\ \leq\ \Delta^{AB}F\ \leq\ W^+_\lambda.
\end{equation}
From \R{I3a} follows 
\begin{equation}\label{I3e}
(W^+_\lambda -W^{AB}_\lambda) P^+_\lambda+(W^-_\lambda -W^{AB}_\lambda) P^-_\lambda\ =\ 0,\quad\longrightarrow\quad
1\ =\ \frac{(W^{AB}_\lambda-W^-_\lambda) P^-_\lambda}{(W^+_\lambda -W^{AB}_\lambda) P^+_\lambda}.
\end{equation}

\section{Jarzynki's Equality\label{JE}}

\subsection{{Two phenomenological lemmata}}

Approaching Jarzynski's equality, an axiom is needed which
postulates suitable properties of the exponential mean process work $M_\lambda$
which is up to now protocol dependent according to \R{bJ10g}. For more physical
elucidation, we formulate this axiom in two steps by two lemmata, so to say as
auxiliary axioms.
Although the exponential mean process work $M_\lambda$ is smaller than
$W^{AB}_\lambda$ according to \R{cJ10}, we 
demand that it satifies the
Second Law \R{aJ4} like the phenomenological process work:\nolinebreak
\vspace{.3cm}\newline
$\blacksquare${\bf Lemma I:}
\begin{equation}\label{dJ10}
\Delta^{AB}F\ \st{\td}{\leq} \ M_\lambda.
\hspace{4.3cm}\blacksquare\hspace{-5cm}
\vspace{.3cm}\end{equation}
Lemma I together with the inequalities \R{cJ10}, \R{hJ10}, and \R{nJ10c} can be summarized as
\begin{equation}\label{dJ10z}
 M^-_\lambda\ <\
\Delta^{AB}F\ \leq \ M_\lambda\ 
\left\{
\begin{array}{l}
\leq\ W^{AB}_\lambda\\
\leq\ M^+_\lambda\\
\end{array}
\right.
\end{equation}
or, equivalently,
\begin{equation}\label{dJ10y}
\exp{(-\beta M^-_\lambda)}\ >\ \exp{(-\beta\Delta^{AB}F)}\ \geq\ 
\exp{(-\beta M_\lambda)}\  
\left\{
\begin{array}{l}
\geq\ \exp{(-\beta W^{AB}_\lambda)}\\
\geq\ \exp{(-\beta M^+_\lambda)}\\
\end{array}
\right..
\vspace{.3cm}\end{equation}
Because Lemma I is demanded for all protocols -- also for reversible ones -- we obtain from \R{bJ10g}$_1$ in comparison 
with \R{dJ10} 
\begin{equation}\label{dJ10a}
M_\lambda^{ABrev}\ =\ \Delta^{AB}F,
\end{equation}
and \R{bJ10e} results in
\begin{equation}\label{dJ10b}
\int p^{rev}_\lambda(x)\exp(-\beta x)dx\ =\ \exp(-\beta\Delta^{AB}F).
\end{equation}
That is to say, Lemma I implies the Jarzynski equality
~(\ref{eq Jarzynski equality}) for reversible protocols.
Taking the second inequality of \R{dJ10y} into account, \R{nJ10a} results in
\begin{equation}\label{dJ10t}
P^-_\lambda\ \leq\ \frac{\exp(-\beta\Delta^{AB}F)-\exp(-\beta M^+_\lambda)}
{\exp(-\beta M^-_\lambda)-\exp(-\beta M^+_\lambda)},
\end{equation}
that means,
Lemma I gives also  a constraint on the integrated probability $P_\lambda^-$ of the
non-regular processes, an inequality which we need later on.
\vspace{.3cm}\newline
Now, to extend the validity of the Jarzynski equality to arbitrary protocols we introduce
a second 
Lemma, which states that the non-regular admixture \R{nJ10a} should have an influence as great as possible by choosing 
$M_\lambda$ independently of the special protocol. Hence, we demand
\vspace{.3cm}\newline
$\blacksquare${\bf Lemma II:}
\begin{eqnarray}\nonumber
\Big(P^-_\lambda\rightarrow\mbox{max, for all protocols}\Big)
\longrightarrow\hspace{5.5cm}
\\ \label{J11}\longrightarrow
\Big(\exp(-\beta M_\lambda)\rightarrow\mbox{max}\Big)
\longrightarrow\Big(M_\lambda\ \st{\td}{=}\ \Delta^{AB}F\Big).
\hspace{1cm}
\blacksquare\hspace{-.7cm}
\vspace{.3cm}\end{eqnarray}
Note that Lemma II implies Lemma I, but we found it intuitive to start with Lemma I separately. 
Furthermore, the second inequality of \R{dJ10y} and the inequality \R{dJ10t} change into equations by Lemma II. 
Now, multiplication of \R{bJ10}$_1$ with $\exp(\beta M_\lambda)$ and taking
Lemma II into account results in a 
phenomenological vindication of Jarzynski's equality
\begin{equation}\label{J13}
 \boxed{
  \{{\cal J}_\lambda\}:\hspace{1cm} \int p_\lambda(x)\exp\Big(-\beta (x-\Delta^{AB}F)\Big)dx\ =\ 1
 }
\end{equation}
including \R{dJ10b} which can be derived without using Lemma II. 
\vspace{.3cm}\newline
Finally, taking Jarzynski's equality into account, from \R{nJ10} and \R{nJ10a}
follows with \R{J11} for the admixtures of the regular and non-regular processes
\begin{eqnarray}\label{I2}
P^+_\lambda\ =\ 
\frac{\exp(-\beta M^-_\lambda)-\exp(-\beta \Delta^{AB}F)}
{\exp(-\beta M^-_\lambda)-\exp(-\beta M^+_\lambda)},
\\ \label{I3}
P^-_\lambda\ =\ \frac{\exp(-\beta \Delta^{AB}F)-\exp(-\beta M^+_\lambda)}
{\exp(-\beta M^-_\lambda)-\exp(-\beta M^+_\lambda)},
\end{eqnarray}
resulting in 
\begin{equation}\label{aJ11c}
\frac{P^-_\lambda}{P^+_\lambda}\ =\ 
\frac{1-\exp\Big(-\beta (M^+_\lambda-\Delta^{AB}F)\Big)}
{\exp\Big(+\beta (\Delta^{AB}F-M^-_\lambda)\Big)-1}.
\end{equation}
Note that the restriction on the probability densities $p_\lambda({\cal W^{AB}_\lambda})$ by the 
phenomenolo\-gi\-cal Lemmata I and II ge\-ne\-rating Jarzynski's equality can be tested by experimental investigation 
according to our basic assumption that these probability densities are experimentally given in the view of Non-equilibrium Thermodynamics.
Especially, testing the restrictions on $P_\lambda^\pm$ might require 
much less statistics than the validation of the Jarzysnki equality itself for which it is extremely important to 
sample the very rare events where the dissipated work is much smaller than the free energy difference 
\cite{JarzynskiPRE2006, HalpernJarzynskiArXiv2016}.

\subsection{The Second Basic Axiom\label{SBA}}
\label{sec the second basic axiom}

Because $\Delta^{AB}F$ is a constant belonging to all Jarzynski processes
between $A$ and $B$, we obtain from \R{J13} for two different protocols
$\lambda(t)$ and $\mu(t)$ of $\{{\cal J}_\lambda\}$
\begin{equation}\label{J11c}
\int p_\lambda(x)\exp(-\beta x)dx\ =\ \int p_\mu(x)\exp(-\beta x)dx\ =\
\exp(-\beta \Delta^{AB}F).
\end{equation}
That means, the expectation value of the exponential process work 
$\exp(-\beta{\cal W}^{AB}_\lambda)$
is protocol-independent, and all protocols of the Jarzynski process class have
to satisfy Jarzynski's equality, a fact which restricts the possible probability
densities. Consequently, Jarzynski's equality is an object of experimentally testing
because in Non-equilibrium Thermodynamics we do not start out with special
given proba\-bi\-li\-ty densities $p_\lambda({\cal W^{AB}_\lambda})$.
\vspace{.3cm}\newline
Jarzynski's equality is here established by the two phenomenological lemmata
\R{dJ10} and \R{J11}, whereby the second one includes the first. The two-step
procedure is chosen because of the more evident physical interpretation. The
main result of Jarzynski's equality is that the mean value of the exponential 
process work is protocol-independent according to Lemma II \R{J11}$_3$.
This fact can be used for replacing the two phenomenological lemmata by another
basic axiom:\vspace{.3cm}\newline
$\blacksquare${\bf Second Basic Axiom:}\vspace{-.2cm}
\begin{center}
\fbox{
\parbox[t]{6.3cm}{The mean value of the exponential 
process work is protocol-independent.}}
\end{center}
\hfill$\blacksquare$
\newline
Using this axiom, \R{J11} follows immediately from \R{bJ10g}, because 
protocol independence of the mean value of the exponential  process work means
$M^{rev}_\lambda = M^{nst}_\lambda$. This more formal axiom allows to establish
Jarzynski's equality with out use of Lemmata I and II which can be regarded as physical
interpretation behind the Second Basic Axiom.
\vspace{.3cm}\newline
Jarzynski's equality, derived in the framework of Stochastic Thermodynamics
\cite{1,SEIF}, is an {\em integral fluctuation relation} with regard to the
Jarzynski process class. Whatever its derivation in Stochastic Thermodynamics
may be, from the point of view of Non-equilibrium Thermodynamics, Jarzynski's
equality can be phenomenologically established by the Second Basic Axiom.
The procedure for implementing stochastic processes into Non-equilibrium Thermodynamics is totally different 
from that used in Stochastic Thermodynamics because we neither make use of any underlying equation of motion nor any specific Hamiltonian in our framework.
In Non-equilibrium Thermodynamics non-regular processes 
appear instead of reversed processes with the difference that non-regular
processes are measurable contributions to the non-regular admixture.
The Second Basic Axiom which allows to establish Jarzynski's equality is a
phenomenological statement on the protocol-independence of the mean values
of the exponential process works.

\section{Some Results}
\subsection{Dissipation and non-stochasticity}

Jarzynski's equation allows to express the dissipation. Starting out with
\R{J13} and \R{J10}
\begin{equation}\label{J11d}
\int p_\lambda(x)\exp(-\beta x)dx\ =\ \exp(-\beta \Delta^{AB}F)\ \geq\
\exp(-\beta W^{AB}_\lambda),
\end{equation}
we obtain
\begin{equation}\label{J11e}
\int p_\lambda(x)\exp\Big(-\beta (x-W^{AB}_\lambda)\Big)dx\ =\
\exp\Big(+\beta (W^{AB}_\lambda-\Delta^{AB}F)\Big)\ \geq\ 1,
\end{equation}
and the dissipation is
\begin{equation}\label{J11f}
D^{AB}_\lambda\ :=\ \beta (W^{AB}_\lambda-\Delta^{AB}F)\ =\
\ln\int p_\lambda(x)\exp\Big(-\beta (x-W^{AB}_\lambda)\Big)dx\ \geq\ 0.
\end{equation}
The phenomenological process work $W^{AB}_\lambda$ is given by \R{J9}$_2$.
\vspace{.3cm}\newline
A further result due to Jarzynski's equality is obtained for non-stochastic
processes: taking \R{J11}$_3$ into account, \R{cJ10c} yields
\begin{equation}\label{I1}
\Delta^{AB}F\ =\ M^{nst}_\lambda\ =\ W^{ABnst}_\lambda\ \geq\ \Delta^{AB}F
\ \longrightarrow\ \Delta^{AB}F\ =\ W^{ABnst}_\lambda,
\end{equation}
that means, non-stochastic processes are always reversible, if Jarzynski's
equality holds:
\begin{equation}\label{J11g}
\mbox{Jarzynski's equality}\ \longrightarrow
\left\{
\begin{array}{ccc}
\mbox{non-stochastic} & \longrightarrow & \mbox{reversible} \\
\mbox{stochastic} & \longleftarrow & \mbox{irreversible} \\
\end{array}
\right.
\vspace{.3cm}\end{equation}
Because reversible "processes" are defined as trajectories in the equilibrium 
sub-space \C{WM1,WM2,WM3}, they are idealized objects which do not exist in nature.
Nevertheless, the reversible processes have to be included into the theoretical
framework because they belong to it as a closure of the theory.
According to \R{J11g}, all irreversible processes 
create stochasticity in the sense that the work distribution is different from a delta 
distribution, but why was this not recognized so far within the framework of 
phenomenological Non-equilibrium Thermodynamics? This is due to the fact that most 
experiments were carried out on macroscopic systems where the number of repetitions 
of the experiment as well as the measurement device is not sensible enough to 
discriminate between different work values for a given protocol. Hence, the 
different process works appear as being equal according to \R{J4} and \R{J5}
\begin{equation}\label{I1a} 
{\cal J}_\lambda\!\!\!\int_A^B \Big[{\cal L}({\lambda,\Theta})-
L({\lambda,\Theta})\Big]\st{\td}{\lambda}(t)dt\ \approx\ 0.
\end{equation}
Consequently, conventional Non-equilibrium
Thermodynamics is a special case of Stochastic Thermodynamics, if Jarzynski's equality
holds and the stochasticity of the irreversible processes is ignored. Another possibility
to ignore stochasticity is to remove the non-regular processes from the theoretical
concept, discussed in the next section.

\subsection{Stochasticity without "violations" ?}

Taking lemma II \R{J11} into account, Jarzynski's equality writes according to \R{fJ10}
\begin{equation}\label{E1}
\int_{x\geq\Delta} p_\lambda(x)\exp(-\beta x)dx
+\int_{x <\Delta} p_\lambda(x)\exp(-\beta x)dx\
=\ \exp(-\beta\Delta^{AB}F).
\end{equation}
Here, the $(x\geq\Delta)$-terms belong to the regular processes.
According to \R{iJ10}$_2$, the non-regular admixture includes all experiments
belonging to protocols $\lambda(t)$ whose process work $x$ is smaller than
the difference $\Delta^{AB}F$ of the free energy. These expe\-ri\-ments are
some times denoted as "violations" of the Second Law. This expression should 
be used with care because the phenomenological Second Law is not a statement
valid for stochastic variables.
\vspace{.3cm}\newline
We now investigate the consequences, if the non-regular processes are eliminated
and only the regular processes are considered. Thus, we set
\begin{equation}\label{E2}
x\ <\ \Delta^{AB}F:\quad p_\lambda(x)\ \doteq\ 0\quad\longrightarrow
\quad \int_{x\geq\Delta} p_\lambda(x)dx\ =\ 1,
\end{equation}
and a naive application of Jarzynski's equality \R{E1} becomes
\begin{equation}\label{E3}
\int_{x\geq\Delta} p_\lambda(x)\Big(\exp(-\beta x)-
\exp(-\beta\Delta^{AB}F)\Big)dx\ =\ 0.
\end{equation}
Because the big bracket in \R{E3} is negative, the probability density is
\begin{equation}\label{E4}
x\ \geq\ \Delta^{AB}F:\quad p_\lambda(x)\ =\ \delta(x-\Delta^{AB}F),
\end{equation}
that means: if Jarzynski's equality holds and 
if the non-regular admixture va\-ni\-shes, the stochastic process
works $x={\cal W}^{AB}_{\lambda}$ have for all protocols the same value 
$\Delta^{AB}F$. Consequently, the regular processes are non-stochastic
according to \R{cJ10a} and reversible according to \R{J11g}. Thus, we proved the
following statement:
\begin{center}\fbox{
\parbox{10.5cm}{If Jarzynski's equality holds and the non-regular
admixture vanishes, all regular processes are non-stochastic and reversible.}}
\end{center}
or shorter in other words: irreversible processes generate stochasticity and
no stochas\-ti\-ci\-ty without non-regular processes.

\subsection{A special family of probability densities}

In principle, there are many different probability distributions 
possible which sa\-tis\-fy the Jarzynski equality. But of course, these probability distributions
are not arbitrary because they have to satify Jarzynski's equality as a constraint.
We now are going to consider a special, but characteristic family for which the
non-regular processes are much less frequent than the regular ones.
For this pupose we start out with Jarzynski's equality \R{J13} written down in the
special decomposition into regular and non-regular processes
\begin{eqnarray}\nonumber
\int_{x\geq\Delta} p_\lambda(x)\exp\Big(-\beta (x-\Delta^{AB}F)\Big)dx +
\int_{x <\Delta} p_\lambda(x)\exp\Big(-\beta (x-\Delta^{AB}F)\Big)dx\ =\
\\ \label{aI3}
=\ 1\ =\ \int_{x\geq\Delta} p_\lambda(x)dx +
\int_{x <\Delta} p_\lambda(x)dx,\hspace{.4cm}
\end{eqnarray}
resulting in
\begin{eqnarray}\nonumber
\int_{x\geq\Delta} p_\lambda(x)
\Big[\exp\Big(-\beta (x-\Delta^{AB}F)\Big)-1\Big]dx\ =\hspace{3cm}
\\ \label{bI3}
=\ \int_{x<\Delta} p_\lambda(x)
\Big[1-\exp\Big(-\beta (x-\Delta^{AB}F)\Big)\Big]dx.
\end{eqnarray}
By changing the integral limits: first integral $x=:\Delta^{AB}F+y$, second integral
\newline$x=:\Delta^{AB}F-y$, $y\geq 0$
\begin{eqnarray}\nonumber
\int_0^\infty p_\lambda\Big(\Delta^{AB}F+y\Big)\Big[\exp(-\beta y)-1\Big]dy\ =\hspace{3cm}
\\ \label{cI3}
=\ \int_0^\infty  p_\lambda\Big(\Delta^{AB}F-y\Big)\Big[1-\exp(\beta y)\Big]dy.
\vspace{.3cm}\end{eqnarray}
A special probability density obeying \R{cI3} and consequently also satisfying
Jarzyns\-ki's equality is
\begin{equation}\label{eI3}
 p_\lambda\Big(\Delta^{AB}F-x\Big)\ =\ 
\frac{1-\exp(-\beta x)}{\exp(\beta x)-1}p_\lambda\Big(\Delta^{AB}F+x\Big),\quad
x>0.
\end{equation}
Because
\begin{equation}\label{fI3}
\frac{1-\exp(-\beta x)}{\exp(\beta x)-1}\ =\ \exp(-\beta x),\quad x>0,
\end{equation}
we obtain for the probability density \R{eI3}
\begin{equation}\label{gI3}
 p_\lambda\Big(\Delta^{AB}F-x\Big)\ =\ \exp(-\beta x)
p_\lambda\Big(\Delta^{AB}F+x\Big),\quad
x>0,
\end{equation}
showing that the probability density belonging to the non-regular 
processes is exponentially smaller than that belonging to the
regular ones. This special chosen case motivates to formulate a further
axiom in the next section which holds generally for all probability
densities and all protocols of the Jarzynski process class and not only for the special
chosen case.
\vspace{.3cm}\newline
The result \R{gI3}is here derived by a phenomenological procedure.
As Crooks found out in Stochastic Thermodynamics
~\cite{CrooksJSP1998, CrooksPRE1999, CrooksPRE2000}, the distribution
belonging to the non-regular processes can be linked to the experiment 
by considering the \emph{inverse} protocol of $\lambda(t)$, defined by
$\lambda^\dagger(t) \equiv \lambda(\tau-t)$, $t\in[0,\tau]$, and the inversion of
any magnetic field and rotation. \R{gI3} is then known as a special case 
of the so-called \emph{detailed fluctuation theorem}~\cite{SEIF}.

\section{The Non-regular Admixture}

In the last section,
a special class of probability densities was distinguished by an arbitrary choice
which results in the fact that the probability density of the non-regular processes
is smaller than that of the regular processes. This statement
has to be generalized for other process
quantities presupposing Jarzynski's equality.

\subsection{The third axiom}

Inspired by the last section, we postulate the third axiom:
\vspace{.3cm}\newline
$\blacksquare${\bf Third Axiom:} For all Jarzynski processes belonging
to an arbitrary chosen protocol $\lambda (t)$, the fraction 
of the non-regular processes  is not greater than that of the regular ones
\begin{equation}\label{I4}
P^-_\lambda\ \st{\td}{\leq}\ P^+_\lambda
\quad\longrightarrow\quad
P^-_\lambda \leq \frac{1}{2},\quad P^+_\lambda\geq\frac{1}{2}.
\hspace{2.4cm}\blacksquare\hspace{-2.4cm}
\vspace{.3cm}\end{equation}
This axiom goes beyond the Jarzynski equality or the Crooks 
fluctuation theorem because we posutlate it for every protocol of the Jarzynski process
class.
\vspace{.3cm}\newline
A first consequence of \R{I4} is according to \R{I3e}
\begin{equation}\label{Z4}
W^{AB}_\lambda-W^-_\lambda\ \geq\ W^+_\lambda -W^{AB}_\lambda
\quad\longrightarrow\quad 2W^{AB}_\lambda\ \geq\  W^+_\lambda + W^-_\lambda.
\end{equation}
This result \R{Z4}$_1$ is pretty clear: the difference between the
phenomenological process work and the  mean value of the process works of
the non-regular admixture is not smaller than that for the regular processes.
This statement is so evident that it could serve as an axiom instead of \R{I4}.
We now go back to the exponential mean values of the process work. 
\vspace{.3cm}\newline
Introducing the abbreviations according to \R{hJ10}$_3$
\begin{equation}\label{Z4a}
a\ :=\ \beta(M^+_\lambda - \Delta^{AB}F)\ \geq\ 0,\qquad
b\ :=\ \beta(\Delta^{AB}F -M^-_\lambda)\ \geq\ 0,
\end{equation}
we obtain from \R{aJ11c} and \R{I4}$_1$                          
\begin{eqnarray}\nonumber
\frac{P^-}{P^+}\ =\ \frac{1-\exp(-a)}{\exp b -1}\ \leq\ 1&\longrightarrow&
2\ \leq\ \exp b + \exp(-a)
\\ \label{Z4b}
&\longrightarrow& \ln\Big(2-\exp(-a)\Big)\ \leq\ b,
\end{eqnarray}
an inequality connecting $M^+_\lambda$, $M^-_\lambda$ and $ \Delta^{AB}F$
according to the third axiom. Because of
\begin{equation}\label{1Z4a}
2-\exp(-a)\ \leq\ \exp a
\end{equation}
\R{Z4b} results in
\begin{equation}\label{2Z4a}
\ln\Big(2-\exp(-a)\Big)\ =\ b_{min}(a)\ \leq\ a.
\end{equation}
Consequently, if $a$ is given, two kinds of $b$ ($b^-$ and $b^+$)
are possible satisfying \R{Z4b} in accordance with the third axiom \R{I4}$_1$  
\begin{equation}\label{3Z4a}
b_{min}(a)\ \leq\ b^-\ \leq\ a\ \leq\ b^+.
\end{equation}
Inserting the abbreviations \R{Z4a}, we obtain inequalities which have
$M^+_\lambda$, $M^-_\lambda$ and $ \Delta^{AB}F$ to satisfy so that
the third axiom \R{I4}$_1$ is true
\begin{equation}\label{4Z4a}
\ln\Big(2-\exp\Big(-\beta(M^+_\lambda-\Delta^{AB}F)\Big)\Big)\ \leq\
\left\{\begin{array}{l}
\beta(\Delta^{AB}F-M^-_\lambda)\leq\beta(M^+_\lambda-\Delta^{AB}F)
\\
\beta(M^+_\lambda-\Delta^{AB}F)\leq\beta(\Delta^{AB}F-M^-_\lambda)\\
\end{array}\right.
\end{equation}
In any case, the third axiom enforces
\begin{equation}\label{5Z4a}
\fbox{$
\ln\Big[2-\exp\Big(-\beta(M^+_\lambda-\Delta^{AB}F)\Big)\Big]\ \leq\
\beta(\Delta^{AB}F-M^-_\lambda)
$},
\end{equation}
an inequality which can be tested by experiments.
\vspace{.3cm}\newline
The third axiom allows an estimation of the non-regular admixture. Starting
out with \R{Z4b}$_1$, a $y$ exists
\begin{equation}\label{6Z4a}
\frac{P^-}{P^+}\ =\ \frac{1-\exp(-a)}{\exp b -1}\ \leq\ \exp(-y)\ \leq\ 1, \qquad
y\ \geq\ 0,
\end{equation}
and we obtain
\begin{equation}\label{7Z4a}
\ln\frac{\exp b -1}{1-\exp(-a)}\ \geq\ y. 
\end{equation}
We now apply the estimation
\begin{equation}\label{8Z4a}
\ln x\ \geq\ 2\frac{x-1}{x+1}, \qquad x\ >\ 0.
\end{equation}
Inserting
\begin{eqnarray}\label{7aZ4a}
x-1\ \longrightarrow\quad
\frac{\exp b -1}{1-\exp(-a)} -1 &=& \frac{\exp b+\exp(-a)-2}{1-\exp(-a)}
\\ \label{7bZ4a}
x+1\ \longrightarrow\quad
\frac{\exp b -1}{1-\exp(-a)}+1 &=& \frac{\exp b-\exp(-a)}{1-\exp(-a)}
\end{eqnarray}
into \R{8Z4a}, we obtain by taking \R{Z4b}$_2$ into consideration
\begin{equation}\label{9Z4a}
\ln\frac{\exp b -1}{1-\exp(-a)}\ \geq\
2\frac{\exp b+\exp(-a)-2}{\exp b-\exp(-a)}\ =:\ y\ \geq\ 0.
\end{equation}
Consequently, \R{6Z4a}$_1$ results in
\begin{eqnarray}\label{10Z4a}
\frac{P^-}{P^+} &\leq&
\exp \Big[2\frac{2-\exp b-\exp(-a)}{\exp b-\exp(-a)}\Big]\ =
\\ \label{11Z4a}
&=& \exp \Big[2\frac{2-\exp\Big(\beta(\Delta^{AB}F-M^-_\lambda)\Big)
-\exp\Big(-\beta(M^+_\lambda-\Delta^{AB}F)\Big)}
{\exp\Big(\beta(\Delta^{AB}F-M^-_\lambda)\Big) -\exp\Big(-\beta(M^+_\lambda-\Delta^{AB}F)\Big)}\Big],\hspace{1.2cm}
\end{eqnarray}
if \R{Z4a} is inserted.
\vspace{.3cm}\newline
We now investigate the consequences of the third axiom \R{I4} for equilibrium
and in the case of reversible protocols.

\subsection{Equilibrium: the fourth axiom}

Equilibrium and reversible processes are two different concepts which should be
distinguished properly: whereas reversible "processes" as trajectories in the 
equilibrium sub-space \C{WM2} are defined by \R{cJ10d},
equilibrium is defined by {\em equilibrium conditions} which by
definition are also valid for reversible processes. 
In more detail: taking into account that \R{I3e} is also valid 
for reversible processes, we obtain by use of \R{cJ10d}
\begin{equation}\label{Z4j}
1\ =\ \frac{(W^{AB}_{\lambda rev}-W^-_{\lambda rev}) P^-_{\lambda rev}}
{(W^+_{\lambda rev} -W^{AB}_{\lambda rev}) P^+_{\lambda rev}}\ =\
\frac{(\Delta^{AB}F-W^-_{\lambda rev}) P^-_{\lambda rev}}
{(W^+_{\lambda rev} -\Delta^{AB}F) P^+_{\lambda rev}}.
\vspace{.3cm}\end{equation}
We now have to postulate the equilibrium condition in agreement with the
definition of reversible processes:
\vspace{.3cm}\newline
$\blacksquare${\bf Fourth Axiom (Equilibrium):}
In equilibrium {\bf--}and consequently also for reversible processes{\bf--} 
regular and non-regular processes are equally frequent\footnote{In Stochastic
Thermodynamics, this axiom runs: Forward and backward path probabilities are
equal in equilibrium \C{Ford}}: according to \R{Z4j},
we obtain  
\begin{equation}\label{Z4l}
\frac{P^-_{\lambda eq}}{P^+_{\lambda eq}}\ \st{\td}{=}\ 1
\ \longleftrightarrow\
\frac{\Delta^{AB}F-W^-_{\lambda rev}}{W^+_{\lambda rev} -\Delta^{AB}F}\
=\ 1\ \longleftrightarrow\
2\Delta^{AB}F\ =\ W^+_{\lambda rev}+W^-_{\lambda rev}.
\hspace{.2cm}\blacksquare
\vspace{.3cm}\end{equation}
The fourth axiom \R{Z4l}$_1$ points out that in equilibrium the regular
processes are equalized by the non-regular ones. This corresponds to the assumption
in Stochastic Thermodynamics that in equilibrium the global detailed balance is
satisfied and that probability fluxes are balanced and no net currents appear.
 Here, the non-regular processes come into consideration by splitting the process
integrals according to \R{fJ10} and \R{iJ10}. 
\vspace{.3cm}\newline
According to the fourth axiom \R{Z4l}$_1$ and \R{Z4b}, we obtain by taking
\R{Z4a} and \R{2Z4a} into account
\begin{eqnarray}\label{Z5}
\ln\Big[2-\exp(-a_{rev})\Big] &=& b_{rev}\ \leq\ a_{rev},
\\ \label{Z6}
\ln\Big[2-\exp\Big(-\beta(M^+_{\lambda rev}-\Delta^{AB}F)\Big)\Big] &=&
\beta(\Delta^{AB}F-M^-_{\lambda rev}),
\\ \label{Z7}
\frac{b_{rev}}{\beta}\ =\ 
\Delta^{AB}F -M^-_{\lambda rev}\
&\leq& M^+_{\lambda rev} - \Delta^{AB}F\ =\ \frac{a_{rev}}{\beta},
\end{eqnarray}
and from \R{Z4l}$_3$ and \R{Z7} follows
\begin{equation}\label{Z4p}
2\Delta^{AB}F\ =\ W^+_{\lambda rev}+W^-_{\lambda rev}\ \leq\
M^+_{\lambda rev} + M^-_{\lambda rev}.
\vspace{.3cm}\end{equation}
The fourth axiom \R{Z4l}$_1$ enforces the equality in \R{5Z4a}
\begin{equation}\label{Z8}
\ln\Big[2-\exp\Big(-\beta(M^+_{\lambda rev}-\Delta^{AB}F)\Big)\Big]\ =\
\beta(\Delta^{AB}F-M^-_{\lambda rev})
\end{equation}
which also can be experimentally tested, if the corresponding protocol is reversible,
that means, if the protocol of the Jarzynski process \R{J1} to \R{J3} is suffiently slow $\lambda(\alpha t),\ \alpha\rightarrow 0$, for approximating reversibility.

\section{Summary}

\begin{itemize}

\item Processes of a closed discrete system between to fixed equilibrium states
controlled by a heat reservoir and divided into two parts --the first part with
power exchange caused by fixed initial and fixed final work variables, the se\-cond
one as relaxation by constant work variables to the final equilibrium state--
constitute the Jarzynski process class (\ref{J1}) to (\ref{J3}) and (\ref{J5a}). 

\item Experimental fact is that the process work is different for several identical
Jarzynski processes enforcing to treat the process work as a stochastic variable
whose different values establish the domain of a probability density \R{J5}.

\item The processes of the Jarzynski process class
fall into two branches: regular processes of non-negative process dissipation
(\ref{iJ10})$_1$ and non-regular processes of  negative process dissipation
(\ref{iJ10})$_2$.

\item Although the process dissipation is negative for non-regular
processes, the mean value of all Jarzynski processes --the phenomenological
dissipation-- is not negative (\ref{N32})$_2$.

\item Two phenomenological lemmata\newline
I) the exponential mean process work obeys the Second Law and\newline
II) the non-regular processes have an extent as great as lemma I) allows,\newline  
are the phenomenological back-ground of the basic axiom
\begin{center}\fbox{
\parbox[t]{7.9cm}{
The mean value of the exponential process work is independent of the Jarzynski
process class.}}
\end{center}
This basic axiom implements Jarzynski's inequality straightforward, \R{J11} and \R{J13}.

\item If Jarzynski's equality holds, non-stochastic processes are always reversible.

\item If Jarzynski's equality holds, irreversible processes generate stochasticity.

\item Beyond Jarzynski's equality: the extent of the non-regular processes
is not greater than that of the regular ones \R{I4}$_1$.

\item Beyond Jarzynski's equality: in equilibrium {\bf--}and consequently for
reversible "processes"{\bf--} the fractions of non-regular and regular
processes are equal, \R{Z4l}$_1$.

\end{itemize}

\section{Discussion}

The tools of Stochastic Thermodynamics are based on Statistical 
Mechanics, whereas Non-equilibrium Thermodynamics is a purely phenomenological
theory. Both theories describe phenomenological processes, Stochastic Thermodynamics
by mean values over its statistical back-ground and Non-equilibrium Thermodynamics
by phenomenological Laws. For a special class of processes {\bf--}the Jarzynski process
class{\bf--} an equality, called Jarzynski's equality, was derived in 1997 by a statistical
procedure containing a probability density stemming from the statistical back-ground.
Now the question arises: can Jarzynski's equality be derived phenomenologically,
if Non-equilibrium Thermodynamics is equipped with stochastic processes ?
\vspace{.3cm}\newline
Considering the Jarzynski process class, an experimental fact is that the process
work is a stochastic variable, that means, performing a Jarzynski process identically
repeated, the process work fluctuates, and a probability density is experimentally
generated. What are now the phenomenological axioms which this probability
density has to obey, so that Jarzynski's equality is valid ? The answer {\bf--}based
on two phenomenological axioms{\bf--} is easy: the mean value of the stochastic
exponential process work is the same for all Jarzynski processes.
\vspace{.3cm}\newline
The Jarzynski processes which are identically repeated can be split into regular
and non-regular ones. By definition, the process dissipation of the non-regular
processes is smaller than the corresponding difference of the free energy. 
By contrast,
the process dissipation of the regular processes is not smaller than the free energy
difference as it is valid for all non-stochastic processes.
The non-regular processes are sometimes confusingly called 
``violations'' of the second law of thermodynamics. 
\vspace{.3cm}\newline
Finally statement: Non-equilibrium Thermodynamics of the Jarzynski process class
can be extended by stochastic processes satisfying Jarzynski's equality which obeys
phenomenological axioms.
\vspace{.5cm}\newline
{\bf Acknowledgement} The co-working of Dr. Philipp Strasberg, Institut f\"ur
Theoretische Physik, TU Berlin, is gratefully acknowleged. Numerous helpful and
sometimes controversal discussions with him makes my phenomenological view
concerning Jarzynski's equality more strict. Several parts of the paper are formulated
by him. All references concerning Stochastic Thermodynamics are generated by
Dr. Strasberg. The paper did not come into being without his help.


\begin{thebibliography}{99}

\bibitem{MU1} Muschik, W.: Survey of some branches of thermodynamics.
 J. Non-Equilib. Thermodyn. {\bf 33}, 165-198 (2008)

\bibitem{SEIF} Seifert, U.: Stochastic thermodynamics, fluctuation theorems
and molecular machines. Rep. Prog. Phys. {\bf 75}, 126001 (58pp)
(2012), sect.3.2.1

\bibitem{1} Jarzynski, C.: Nonequilibrium Equality for Free Energy Differences.
Phys. Rev. Lett. {\bf 78}, 2690-2693 (1997)

\bibitem{SS} Schottky, W., Ulich, H., Wagner, C.: Thermodynamik, 
 Springer, Berlin 1929, Reprint, Springer, Berlin 1973, Erster Teil \S 1

\bibitem{5} Muschik, W.: Empirical Foundation and Axiomatic Treatment of
Non-equilibrium Temperature. Arch. Rational Mech. Anal. {\bf 66}, 379-401 (1977)

\bibitem{6} Muschik, W., Brunk, G.: A Concept of Non-equilbrium Temperature.
Int. J. Engng. Sci. {\bf 15}, 377-389 (1977)

\bibitem{9} Muschik, W.: Contact Quantities and Non-equilibrium Entropy of
Discrete Systems. J. Non-Equilib. Thermodyn. {\bf 34}, 75-92 (2009)

\bibitem{JarzynskiPRE1997} Jarzynski, C.: Equilibrium free-energy differences
from nonequilibrium measurements: 
A master-equation approach. Phys. Rev. E \textbf{56}, 5018 (1997)

\bibitem{CrooksJSP1998} Crooks, G.E.: Nonequilibrium measurements of free energy differences for microscopically 
reversible Markovian systems. J. Stat. Phys. \textbf{90}, 1481-1487 (1998)

\bibitem{CrooksPRE1999} Crooks, G.E.: Entropy production fluctuation theorem and the nonequilibrium work relation 
for free energy differences. Phys. Rev. E \textbf{60}, 2721 (1999)

\bibitem{CrooksPRE2000} Crooks, G.E.: Path-ensemble averages in systems driven far from equilibrium. Phys. Rev. E \textbf{61}, 2361 (2000)

\bibitem{JarzynskiJSM2004} Jarzynski, C.: Nonequilibrium work theorem for a system strongly coupled to a thermal 
environment. J. Stat. Mech. \textbf{P09005} (2004)

\bibitem{TasakiArXiv2000} Tasaki, H.: Jarzynski relations for quantum systems and some applications. arXiv: cond-mat/0009244 (2000)

\bibitem{MukamelPRL2003} Mukamel, S.: Quantum extension of the Jarzynski relation: Analogy with stochastic dephasing. Phys. Rev. Lett. \textbf{90}, 170604 (2003)

\bibitem{CampisiTalknerHaenggiPRL2009} Campisi, M., Talkner, P., H\"anggi, P.: Fluctuation theorem for arbitrary open quantum systems. Phys. Rev. Lett. \textbf{102}, 210401 (2009).

\bibitem{TrepagnierEtAlPNAS2004} Trepagnier, E.H., Jarzynski, C., Ritort, F., Crooks, G.E., Bustamante, C., Liphardt, J.: Experimental test of Hatano and Sasa's nonequilibrium steady-state equality. Proc. Natl. Acad. Sci. \textbf{101}, 15038-15041 (2004)

\bibitem{CollinEtAlNature2005} Collin, D., Ritort, F., Jarzynski, C., S. B. Smith, S.B., 
Tinoco, I., Bustamante, C.: Verification of the Crooks fluctuation theorem and recovery of RNA folding free energies. Nature \textbf{437}, 231-234 (2005)

\bibitem{BustamanteLiphardtRitortPhysToday2005} Bustamante, C., Liphardt, J., 
Ritort, F.: The nonequilibrium thermodynamics of small systems. Phys. Today \textbf{58}, 43-48 (2005)

\bibitem{JarzynskiReview2011} Jarzynski, C.: Equalities and inequalities: irreversibility and the second law of thermodynamics at the nanoscale. Annu. Rev. Condens. Matter Phys. \textbf{2}, 329-351 (2011)

\bibitem{2} Muschik, W.: Aspects of non-equilibrium thermodynamics,
Six Lectures on Fundamentals and Methods. World Scientific, Singapore 1990, sect.1.2

\bibitem{10} Muschik, W.: Fundamentals of Non-equilibrium Thermodynamics
in: Muschik, W. (Ed.): Non-equilibrium Thermodynamics with Application to
Solids.
CISM Courses and Lectures No. 336, Springer, Wien 1993, pp. 1 - 63, sect.3.2

\bibitem{8} Muschik, W., Berezovski, A.: Non-equilibrium contact quantities
and compound defiency at interfaces between discrete systems.
Proc. Estonian Acad. Sci. Phys. Math. {\bf 56}, 133-145 (2007)

\bibitem{GM} de Groot, S.R., Mazur, P.: Non-Equilibrium Thermodynamics.
North-Holland, Leyden 1963, sect.III.1

\bibitem{GaveauSchulmanPLA1997} B. Gaveau and L. S. Schulman, \emph{A general framework for non-equilibrium phenomena: The master equation and its formal consequences},
Phys. Lett. A \textbf{229}, 347-353 (1997).

\bibitem{CrooksPRE2007} G. E. Crooks, \emph{Beyond Boltzmann-Gibbs statistics: Maximum entropy hyperensembles out of equilibrium}, 
Phys. Rev. E \textbf{75}, 041119 (2007). 

\bibitem{EspositoVandenBroeckEPL2011} M. Esposito and C. {Van den Broeck}, 
\emph{Second law and Landauer principle far from equilibrium}, Europhys. Lett. \textbf{95}, 40004 (2011).

\bibitem{OST} Ostoja-Starzewski, M.: Second law violations, continuum
mechanics, and permeability. Continuum Mech. Thermodyn. {\bf 26} 489-501 (2016)

\bibitem{JarzynskiPRE2006} Jarzynski, C.: Rare events and the convergence of exponentially averaged work values. Phys. Rev. E \textbf{73}, 046105 (2006)

\bibitem{HalpernJarzynskiArXiv2016} Halpern, N.Y., Jarzynski, C.: How many trials should you expect to perform to estimate a free-energy difference ?. 
arXiv 1601.02637 (2016)

\bibitem{WM1} Muschik, W.: Existence of non-negative entropy production. In:
Spencer A.J.M. (Ed): Continuum Models of Discrete Systems, , A.A. Balkema,
Rotterdam 1987, Procedings of the 5th International Symposium, Nottingham
14-20 July 1985, pp. 39-45

\bibitem{WM2} Muschik, W.: Fundamentals of Dissipation Inequalities,
I. Discrete Systems. J. Non-Equilib. Thermodyn. {\bf 4}, 277-294 (1979)

\bibitem{WM3} Muschik, W.: Skizze der thermodynamischen Theorien irreversibler
Prozesse. In: Reif, F.: Statistische Physik und Theorie der W\"arme. de Gruyter,
Berlin 1985, pp. 709-741

\bibitem{Ford} Ford, I.: Statistical Physics, An Entropic Approach. Wiley,
2013, sect. 17,2

\end{thebibliography}
\end{document}